\documentclass{pnastwo}

\usepackage[pdftex]{graphicx}

\usepackage{PNAStwoF}
\usepackage{amssymb,amsfonts,amsmath}
\usepackage{framed}

\contributor{Submitted to Proceedings
of the National Academy of Sciences of the United States of America}
\url{www.pnas.org/cgi/doi/10.1073/pnas.1309723111}
\copyrightyear{2014}
\issuedate{Issue Date}
\volume{Volume}
\issuenumber{Issue Number}

\begin{document}

\title{Principles of scientific research team formation and evolution}

\author{
Sta\v{s}a Milojevi\'{c}\affil{1}{School of Informatics and Computing, Indiana University, Bloomington, USA}}

\contributor{Submitted to Proceedings of the National Academy of Sciences
of the United States of America}

\maketitle

\begin{article}

\begin{abstract}
Research teams are the fundamental social unit of science, and yet there is currently no model that describes their basic property: size. In most fields teams have grown significantly in recent decades. We show that this is partly due to the change in the character of team-size distribution. We explain these changes with a comprehensive yet straightforward model of how teams of different sizes emerge and grow. This model accurately reproduces the evolution of empirical team-size distribution over the period of 50 years. The modeling reveals that there are two modes of knowledge production. The first and more fundamental mode employs relatively small, {\it core} teams. Core teams form by a Poisson process and produce a Poisson distribution of team sizes in which larger teams are exceedingly rare. The second mode employs {\it extended} teams, which started as core teams, but subsequently accumulated new members proportional to the past productivity of their members. Given time, this mode gives rise to a power-law tail of large teams (10-1000 members), which features in many fields today. Based on this model we construct an analytical functional form that allows the contribution of different modes of authorship to be determined directly from the data and is applicable to any field. The model also offers a solid foundation for studying other social aspects of science, such as productivity and collaboration.
\end{abstract}


\begin{framed}
\noindent Significance: Science is an activity with far-reaching implications for modern society. Understanding how the social organization of science and its fundamental unit, the research team, forms and evolves is therefore of critical significance. Previous studies uncovered important properties of the internal structure of teams, but little attention has been paid to their most basic property: size. This study fills this gap by presenting a model that successfully explains how team sizes in various fields have evolved over the past half century. This model is based on two principles: (a) smaller (core) teams form according to a Poisson process, and (b) larger (extended) teams begin as core teams but consequently accumulate new members through the process of cumulative advantage based on productivity. 
\end{framed}

\dropcap{C}ontemporary science has undergone major changes in the last half century at all levels: institutional, intellectual, and social, as well as in its relationship with society at large. Science has been changing in response to increasingly complex problems of contemporary society and the inherently challenging nature of unresolved questions, with an expectation to serve as a major driver for economic growth. Consequently, the contemporary science community has adopted a new, problem-driven approach to knowledge production that often blurs the lines between pure and applied, and is more permeable around disciplinary borders, leading to cross-/multi-/inter-/trans-disciplinarity [1]. The major staple of this approach is team effort [2-5]. The increased prominence of scientific teams has recently led to a new research area, {\em science of team science}, which is  ``... centered on examination of the processes by which scientific teams organize, communicate, and conduct research" [6]. If we wish not only to understand contemporary science, but also to create and promote viable science policies, we need to uncover principles that lead to the formation and subsequent evolution of scientific research teams. 

Studies of collaboration in science, and co-authorship as its most visible form, have a long history [7-11]. The collaborative mode of knowledge production is often perceived as being in contrast to the individualistic mode of the past centuries [12, 13]. Previous studies have established that the fraction of co-authored papers has been growing with respect to single-authored papers [5], that in recent decades teams have been growing in size [14], and that inter-institution and international teams are becoming more prevalent [15, 16]. In addition, high-impact research is increasingly attributed to large teams [5, 6], as is research that features more novel combination of ideas [17]. The reasons for an increase in collaborative science have been variously explained as due to the shifts in the types of problems studied [1] and the related need for access to more complex instruments and broader expertise [15, 18, 19]. 

A research team is a group of researchers collaborating in order to produce scientific results, which are primarily communicated in the form of research articles. Researchers who appear as authors on a research article represent a visible and easily quantifiable manifestation of a collaborative, team-science effort. We refer to such a group of authors as an ``article team.'' In this study we focus on one of the most fundamental aspects of team science: {\em article team-size distribution}\footnote{ In the rest of the article we will refer to an article team simply as ``the team."}  and its change/evolution over time. Many studies focused only on the mean or the median sizes of teams, implicitly assuming that the character of the distribution of team sizes does not change. Relatively few studies examined {\em full} team-size distribution, albeit for rather limited data sets [10, 20, 21], with some of them noticing the changing character of this distribution [10]. The goal of the current study is to present a more accurate characterization and go beyond empirical observations to provide a model of scientific research team formation and evolution that leads to the observed team-size distributions. 

Despite a large number of studies of co-authorship and scientific teams, there are few explanatory models. One such exception is Guimera et al.'s model of the self assembly of teams [2], which is based on the role that newcomers and repeated collaborations play in the emergence of large connected communities and the success of team performance. Although their model features team size as a parameter, its values were not predicted by the model but were taken as input from the list of actual publications.  The objective of the current study is to go beyond the internal composition of teams in order to explain the features of team-size distribution and its change over the past half century. Thus, the model we propose in this paper is complementary to Guimera et al.'s efforts. Our model is based on several simple principles that govern team formation and its evolution. The validity of the model is confirmed by constructing simulated team-size distributions that closely match the empirical ones based on 150,000 articles published in the field of astronomy since 1960s. We reveal the existence of two principal modes of knowledge production: one that forms small core teams based on a Poisson process, and the other that leads to large, extended teams that grow gradually on the principle of cumulative advantage.

\section{Empirical team-size distributions}

The significant change in the character of team-size distribution is the key insight underlying the proposed model. Previous studies have shown a marked increase in the {\it mean} team size in recent decades, not only in astronomy [e.g, 2, 22], but in all scientific fields [5]. Specifically, the average team size in astronomy grew from 1.5 in 1961-1965 to 6.7 in 2006-2010 (marked by arrows in Fig.\ 1, which shows, on a log-log scale, team-size distributions in the field of astronomy in two time periods). However, Figure 1 reveals even more: a recent distribution (2006-2010) is not just a scaled-up version of the 1961-1965 distribution shifted towards larger values; it has a profoundly different shape. Most notably, while in 1961-1965 the number of articles with more than five authors was falling precipitously, and no article featured more than eight authors, now there exists an extensive tail of large teams, extending to team sizes of several hundred authors. The tail closely follows the {\it power-law} distribution (red line in Fig.\ 1). The power-law tail is seen in recent team-size distributions of other fields as well [23]. In contrast, the ``original'' 1961-1965 distribution did not feature a power-law tail. Instead, most team sizes were in the vicinity of the mean value. The shape of this original distribution can instead be described with a simple Poisson distribution (blue curve in Fig.\ 1), an observation made in some previous works [10, 20]. Note that the time when the distribution stopped being Poisson would differ from field to field.

\begin{figure}
\centering
\includegraphics[width=0.5\textwidth]{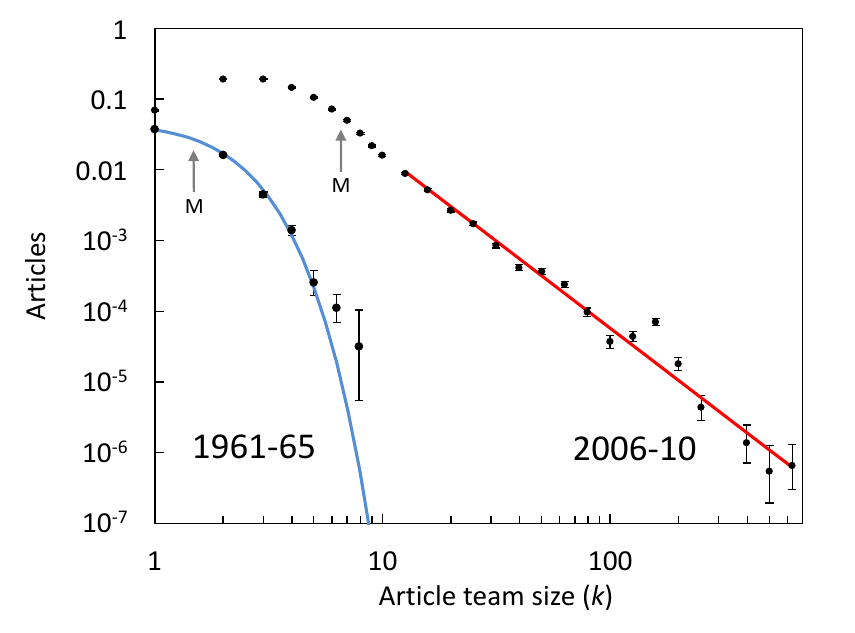}\caption{Distribution of article team sizes in astronomy in two time periods separated by 45 years. The distribution from 1961-1965 is well described by a Poisson distribution (blue curve). This is in contrast to 2006-2010 distribution, which features an extensive power-law tail (red line). Arrows mark the mean values of each distribution. For $k >10$ ($k>5$for 1961-65) the data are binned in intervals of 0.1 decades, thus revealing the behavior far in the tail, where the frequency of articles of a given size is up to million times lower than in the peak. All distributions in this and subsequent figures are normalized to the 2006-2010 distribution in astronomy. Error bars in this and subsequent figures correspond to one standard deviation. The full dataset consists of 154,221 articles published between 1961 and 2010 in four core astronomy journals (listed in SI), which publish the majority of research in this field [24]. Details on data collection are given elsewhere [25].}
\label{fig:emp}
\end{figure}

We interpret the fact that the distribution of team sizes in astronomy in the 1960s is well described as a stochastic variable drawn from a Poisson distribution to mean that {\it initially} the production of a scientific paper used to be governed by a {\em Poisson process} [26, 27]. This is an intuitively sound explanation because many real-world phenomena involving low rates arise from a Poisson process. Examples include pathogen counts [28], highway traffic statistics [29], and even sports scores [30]. Team assembly can be viewed as a low-rate event, because its realization involves few authors out of a very large possible pool of researchers. Poisson rate ($\lambda$) can be interpreted as a characteristic number of authors that are necessary to carry out a study. The actual realization of the process will produce a range of team sizes, distributed according to a Poisson distribution with the mean being this characteristic number. 

In contrast, the dynamics behind the power-law distribution that features in team sizes in recent times is fundamentally different from a simple Poisson process, and instead suggests the operation of a process of {\em cumulative advantage}. Cumulative advantage, also known as the Yule process, and as preferential attachment in the context of network science [31, 32], has been proposed as an explanation for the tails of collaborator and citation distributions [23, 32-38]. Unlike the Poisson process, cumulative advantage is a dynamic process in which the properties of a system depend on its previous state. How did a distribution characterized by a Poisson function evolve into one that follows a power law? Does this evolution imply a change in the mode of the team assembly? Does a Poisson process still operate today? Figure 1 shows that for smaller team sizes ($k < 10$) the power law breaks down, forming instead a ``hook.'' This small-$k$ behavior must not be neglected because the great majority of articles (90\%) are still published in teams with fewer than ten authors. The hook, peaking at teams with two or three authors, may represent a vestige of what was solely the Poisson distribution in the past. This simple assumption is challenged by the fact that no single Poisson distribution can adequately fit the small-$k$ portion of the 2006-10 team-size distribution. Namely, the high ratio of two-author papers to single-author papers in the 2006-10 distribution would require a Poisson distribution with $\lambda = 2 P_{k=2}/ P_{k=1} = 5.5$. Such distribution produces a peak at $k = 5$, which is significantly offset compared to its actual position. Evidently, the full picture involves some additional elements.

In the following section we present a model that combines the aforementioned processes and provides answers to the questions raised in this section, demonstrating that knowledge production occurs in two principal modes.

\section{Model of team formation and evolution}

We next lay out a relatively simple model that incorporates principles of team formation and its evolution.  We produce simulated team-size distributions based on the model and validate them by testing how well they ``predict'' empirical distributions in the field of astronomy. This model is universally applicable to other fields, as will be discussed later.

The model consists of \emph{authors} who write \emph{papers} over time. Each paper has a \emph{lead} author who is responsible for putting together a team and producing a paper. Each lead author is associated with two types of teams: \emph{core} and \emph{extended}. Core teams consist of the lead author and coauthors. Their size is drawn from a Poisson distribution with some rate $\lambda$. If the drawing yields the number one, the core team consists of the lead author alone. We allow $\lambda$, the characteristic size of core teams, to grow with time. Existing authors, when they publish again, retain their original core teams. The probability of publishing by an author who has published previously is 0.8. Unlike core teams, extended teams evolve dynamically. Initially, the extended team has the same members as the core team. However, the extended team is allowed to add new members in proportion to the aggregate productivity of its current members. New extended team members are randomly chosen from core teams of existing members, or from a general pool if no such candidates are available. The cumulative advantage principle that governs the growth of extended teams will mean that teams that initially happen to have more members in their core teams and/or whose members have published more frequently as lead authors, will accrete more new members than the initially smaller and/or less productive teams.\footnote{We have tested several flavors of cumulative advantage and found that the empirical distributions are best reproduced when the growth follows the aggregate productivity of all members as lead authors, rather than their productivity that includes co-authorships.} This process allows some teams to grow very large, beyond the size that can be achieved with a Poisson process. The process is gradual, so very large teams appear only when some time has passed.  It is important that extended teams do not replace core teams; they co-exist, and the lead author can choose to publish with one or the other at any time. This choice is presumably based on the type or complexity of a research problem. In simulation we assume a fixed probability ($p_{\rm ext}=0.3$) for an article to require an extended team. Core and extended teams correspond to traditional and team-oriented modes of knowledge production, respectively.  

\begin{figure}
\centering
\includegraphics[width=0.5\textwidth]{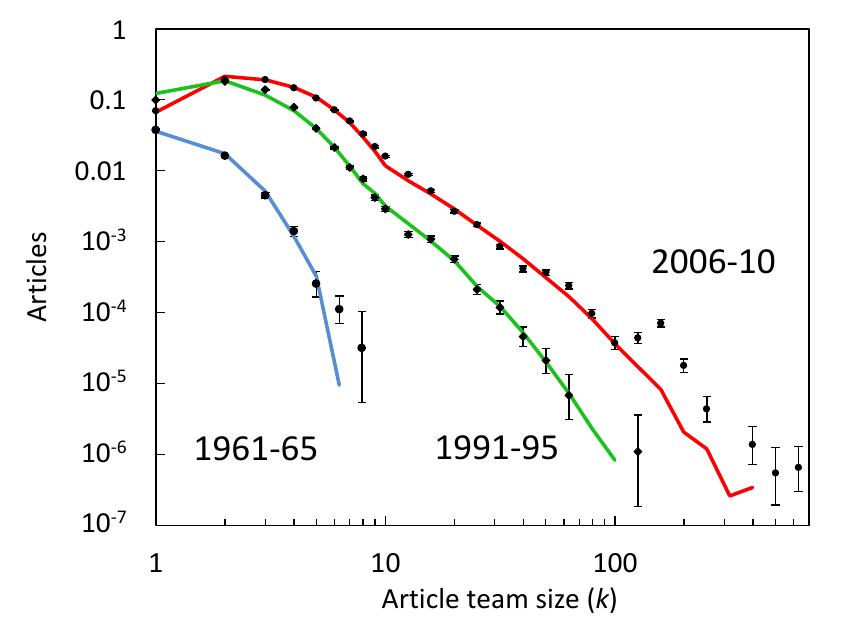}\caption{Comparison between article team-size distributions based on model simulation (values connected by colored lines) and the empirical data (points) for the field of astronomy in three time periods. Our model for the formation and evolution of teams reproduces the observed distributions remarkably well. The model assumes that each lead author forms a core team through a Poisson process. Additionally, extended teams arise from core teams by adding new members in proportion to the productivity of the team. Team growth of productive teams then facilitates further team growth. This process of cumulative advantage leads to the appearance of the power-law component of large teams at later times. In the model, each time a paper is produced the lead author can choose to work with his/her core team or the extended team, thus leading to two main modes of knowledge production. Interestingly, in our simulation the probability of choosing core or extended teams does not need to change over time in order to match the data. Kolmogorov-Smirnov tests were run to formally assess the match between the model and data. For the three time periods shown the maximum deviations are $D=0.11, 0.06, 0.17$, corresponding to $<1\%$ of chance match. All distributions are normalized to the 2006-2010 distribution.}
\label{fig:model}
\end{figure}

\begin{figure}
\centering
\includegraphics[width=0.5\textwidth]{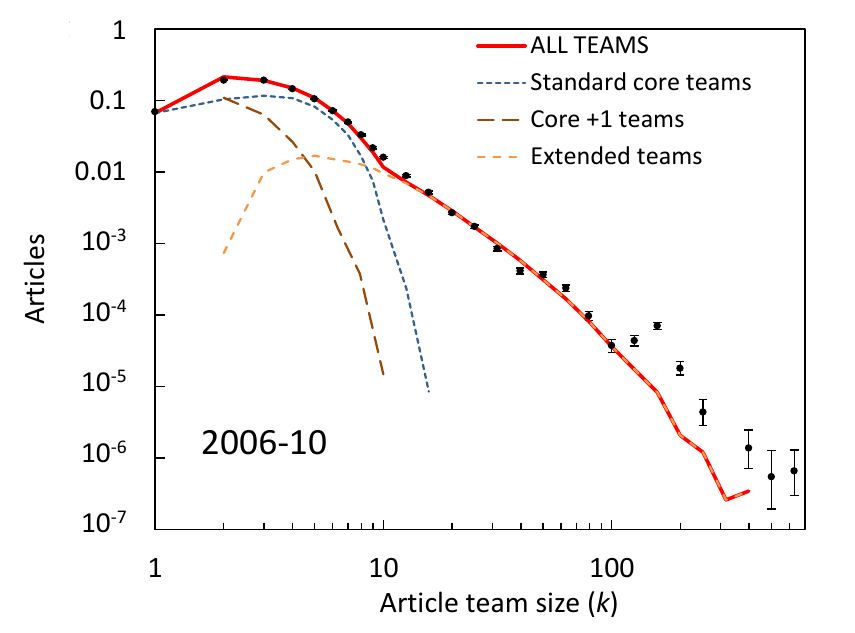}\caption{Distribution of article team sizes according to the generating authorship mode (for astronomy, 2006-2010). A lead author can choose to publish with his/her core team or the extended team. The mode that involves core teams dominates in articles with fewer than ten authors. Furthermore, to accurately reproduce the empirical distribution it is necessary to assume two types of core teams: standard and ``core +1'' teams. The latter type is also drawn from a Poisson distribution, but includes an extra member. The majority of such articles are presumably produced by teams based around student-mentor pairs.}
\label{fig:model_comp}
\end{figure}

We also incorporate several additional elements to this basic outline that brings the model closer to reality. First, the empirical data indicate that in recent times there is an excess of two-author papers over single-author papers, especially from authors who have just started publishing. Apparently, such authors tend not to publish alone, probably because they include their mentors as coauthors. To reproduce such behavior we posit in the model that some fraction of lead authors will form their core teams by adding an additional member to the number drawn from a Poisson distribution. We call such teams ``core +1 teams,'' as opposed to ``standard core teams.'' Furthermore, we assume that repeat publications are more likely from authors who started publishing more recently. Finally, we assume that certain authors retire and their teams are dissolved. However, the process of retirement is not essential to reproduce the empirical team-size distribution.

The model is implemented through a simulation of 154,221 articles, each with a list of ``authors.'' The number of articles is set to match the empirical number of articles published within the field of astronomy in the period 1961-2010. The sequence in which the articles are produced in the simulation allows us to match them to actual publication periods (e.g., articles with sequential numbers 51188 to 69973 correspond to articles published from 1991 to 1995). In Figure 2 we show a compelling match between the real data (dots with error bars) and the predictions of our model (values connected by colored lines) for three time periods (1961-65, 1991-95, and 2006-10). The model correctly reproduces the emergence of the power-law tail and its subsequent increased prominence, as well as the change in the shape of the low-$k$ distribution (the hook), and the shift of the peak from single-author papers to those with two or three authors. The strongest departure of the model from the empirical distribution is the bump in the far tail of the 2006-10 distribution (around $k$ = 200). We have identified this ``excess'' to be due to several papers that were published by a FERMI collaboration [39] over a short period of time. Note, however, that only 0.6\% of all 2006-10 papers were published by teams with more than 100 authors.

In addition to predicting the distribution of team sizes, the model also produces good predictions for other, author-centric distributions. Figure S1 compares model and empirical distributions for article per author (productivity), collaborator per author, and team per author distributions, as well as the trend in the size of the largest connected component. The latter correctly predicts that the giant component forms in the early 1970s. Distributions and trends based on the implementation of Guimera et al.\ team assembly principles [2] are also shown in Figure S1 for comparison (with team sizes supplanted from our model). They yield predictions of similar quality. Collaborator distribution has been the focus of numerous studies [34-38]. Here we follow the usual determination of collaborators based on co-authorship. In the limiting case in which each author appears on only one article (which is true for the majority of authors over time periods of a few years), the collaborator distribution, $F_C$, is related to team-size distribution as: $F_{C}(n) = (n+1)F(n+1)$, where $F$ is the team-size distribution. Therefore, the power-law tail in the collaborator distributions, which has been traditionally explained in the network context as the manifestation of the \emph{preferential attachment} in which authors with many collaborators (``star scientists'' [40]) have a higher probability of acquiring new collaborators (nodes that join the network) may alternatively be interpreted as authors (not necessarily of  ``star" status) belonging to extended teams that grow through the mechanism of cumulative advantage.

Interestingly, the model predicts the empirical distribution quite well (Figure 2), even though we assumed that the propensity to publish with the extended team has remained constant over the 50-year period ($p{\rm ext}=0.3$). This suggests a hypothesis that (at least in astronomy) there always existed a similar proportion of problems that would have required non-individualistic effort, but it took time for such an approach of conducting research to become conspicuous because of the gradual growth of extended teams. 

The model allows us to assess the relative contribution of different modes of authorship. In Figure 3 we separately show the distribution of articles produced by both types of core teams and the extended teams. By definition, ``core +1'' teams and extended teams start at $k = 2$, and therefore single-author papers can only be produced in a standard-core team mode. Two-author teams are almost exclusively the result of core teams with equal shares of standard and ``core +1'' teams. The contribution of ``core +1'' teams drops significantly for three or more authors, which is not surprising if such teams are expected to be primarily composed of student-mentor pairs. Standard core teams dominate as the production mechanism in articles containing up to eight authors; i.e., they make up most of the hook. Extended teams become the dominant mode of production of articles that include ten or more authors, thus they are responsible for the power-law tail of large teams. 

\section{Analytical decomposition of team-size distributions}

Deriving the relative contribution of different types of teams as performed in the previous section and shown in Figure 3 requires a model simulation and is therefore not practical as a means of interpreting empirical distributions. Fortunately, we find (by testing candidate functions using the maximum likelihood method) that the distribution of the articles produced by each of the three types of teams can be approximated by the following functional form equivalents: standard core and ``core +1'' teams are well described by Poisson functions, $F_{P1}(k)$ and $F_{P2}(k)$, while the distribution of articles produced by extended teams is well described by a power-law function with a low-end exponential cutoff, $F_{PL}$.  Therefore, the following analytical function can be fit to the empirical team-size distribution in order to obtain its decomposition:

\begin{equation}
F(k) = F_{P1}(k)+F_{P2}(k)+F_{PL}(k) = \nonumber\\
\end{equation}

\begin{equation}
= \left\{ \begin{array}{ll}
n_{1}\frac {\lambda_{1}^{k} e^{-\lambda_{1}}} {k!}, & k=1\\
n_{1}\frac {\lambda_{1}^{k} e^{-\lambda_{1}}} {k!} + n_{2}\frac {\lambda_{2}^{k-1} e^{-\lambda_{2}}} {(k-1)!} + n_{3}e^{-\beta/(k-1)}k^{-\alpha}, & k>1
\end{array} \right.
\end{equation}

\begin{figure}[h]
\centering
\includegraphics[width=0.5\textwidth]{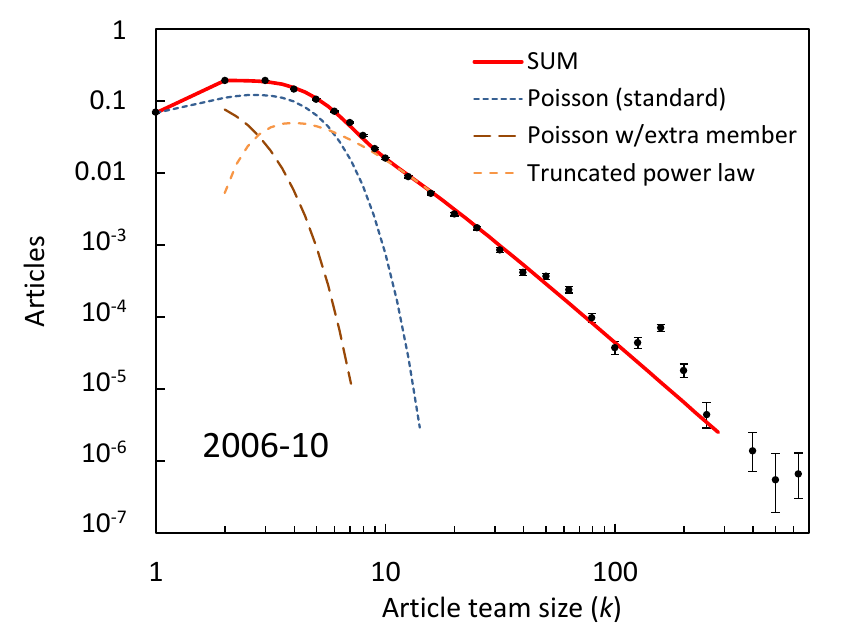}
\caption{Functional decomposition of the empirical article team-size distribution (for astronomy, 2006-2010). Different modes of authorship identified by the model have their functional equivalents, thus allowing the empirical determination of the contribution of each mode to the team-size distribution. Core teams are well fit by Poisson functions, while the extended teams are well fit by an exponentially truncated power-law component. Based on the best-fitting function given in Equation 1, for $k <100$. KS test yields $D=0.05$, corresponding to $<0.1\%$ of chance match.}
\label{fig:emp_comp}
\end{figure}

\begin{figure}[h]
\centering
\includegraphics[width=0.5\textwidth]{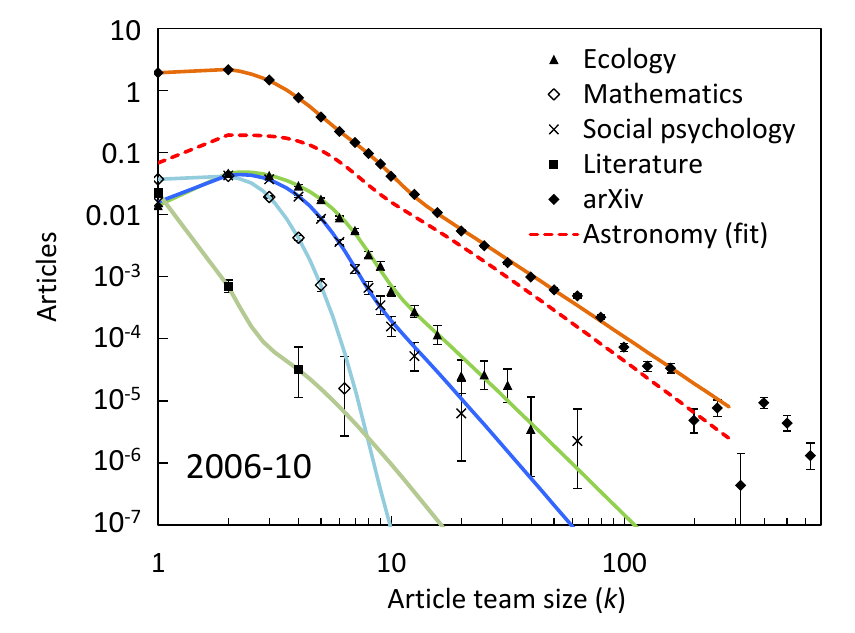}\caption{Functional fits (Equation 1) to the article team-size distribution in the fields of mathematics, ecology, literature, social psychology, and for arXiv (for 2006-2010). All distributions are well fitted by the functional form that is a sum of two Poisson functions and a truncated power law (Equation 1), demonstrating that the proposed analytical description is universal. Distributions are normalized to the 2006-2010 distribution in astronomy, which is also shown with its best-fitting function for comparison (without data points, for clarity). A KS test yields $D=0.06, 0.17, 0.08$ and 0.05 for ecology, mathematics, social psychology, and arXiv respectively, which all correspond to $<0.1\%$ probability of chance match. Literature has too few points for a KS test.}
\label{fig:mat_ecl}
\end{figure}

In the above expression, $\lambda_1$ and $\lambda_2$ are the Poisson rates for $F_{P1}(k)$ and $F_{P2}(k)$, $\alpha$ is the power-law slope, and $\beta$ determines the strength of the exponential truncation. Relative normalization of the three components is given by $n_1$, $n_2$, and $n_3$. This expression features six independent parameters. While other analytical functions can, in principle, also provide a good fit to the overall size distribution, Eq.\ 1 is constructed so that each component corresponds to a respective authorship mode. Furthermore, as shown in Figure S2, removing various components of Eq.\ 1 leads to decreased ability to fit the empirical distribution.

The best-fitting functional form $F(k)$ for the most recent team-size distribution in astronomy is shown in Figure 4.  The fitting was performed using $\chi^2$ minimization. The overall fit is very good and the individual components of Eq.\ 1 match the different modes of authorship, as derived by the model (Figure 3). By integrating these components we find that currently 57\% of articles belong to $F_{P1}$ and can therefore be attributed to standard core teams. Another 12\% are due to ``core +1'' teams ($F_{P2}$), while the remaining 31\% of articles are fit by the truncated power-law component ($F_{PL}$) and can therefore be interpreted as originating from extended teams.

\begin{table*}
\centering
\caption{Characteristics of different fields obtained from analytical decomposition. Meaning of columns is given in caption to Figure 6.}
\begin{tabular}{lllllllllll}
Field & Articles (2006-10) & $\lambda_{P1}$ & $\lambda_{P2}$ & $\alpha_{PL}$ & $f_{P1}$ &  $f_{P2}$ &  $f_{PL}$ & $\mu  _{P}$ & $\mu_{PL}$ & $\mu_{\rm all}$ \\
\hline
Astronomy & 31,473 & 3.25 & 0.67 & 2.8 & 0.52 & 0.11 & 0.37 & 3.21 & 11.20 & 6.14\\
Ecology & 5,420 & 3.23 & 0.83 & 3.8 & 0.62 & 0.13 & 0.25 & 3.20 & 4.58 & 3.54\\
Mathematics & 3,244 & 0.87 & 0.75 & 13.4 & 0.57 & 0.33 & 0.09 & 1.84 & 2.87 & 1.93\\
Social psychology & 4,122 &  2.24 & 1.58 & 4.5 & 0.46 & 0.36 & 0.18 & 2.72 & 3.89 & 2.93\\
Literature & 725 & 0.06 & 0.02 & 5.0 & 0.99 & 0.00 & 0.01 & 1.03 & 3.75 & 1.05\\
arXiv & 235,414 & 1.80 & 4.93 & 2.6 & 0.72 & 0.05 & 0.23 & 2.38 & 6.56 & 3.36\\
\hline
\end{tabular}
\end{table*}

The principles that underlie the proposed model are universal and not field dependent. Only the parameters that specify the rate of growth or the relative strength of the processes will differ from field to field. Consequently the analytical decomposition given by Eq.\ 1 can be applied to other fields. Figure 5 shows the best-fitting functions (Equation 1) to the empirical team-size distributions in: mathematics, ecology, social psychology, literature, and for articles from arXiv, all for the current period (2006-10). Core journals used for these fields are listed in SI. All of the distributions are well described by our model-based functional decomposition. Parameters for the fit and contributions of different authorship modes are given in Table 1. There is much variety. In literature the standard core team mode accounts for nearly the entire output (99\%) with very small teams. Mathematics also features relatively small teams and a steep decline of larger teams. Nevertheless, the functional decomposition implies that 9\% of articles are produced in the extended team mode (see also Fig.\ S3), but these teams are still not much larger than core teams (2.9 vs. 1.8 members on average). Mathematics and social psychology feature the largest share of ``core +1" teams. Team-size distributions for ecology and social psychology both have more prominent power-law tails than mathematics ($\alpha \sim 4$) but they are not yet as extensive as in astronomy ($\alpha \sim 3$). Both fields feature a hook at low $k$ similar to that of astronomy. Finally, articles from arXiv (mostly belonging to the field of physics) have a power-law slope very similar to that of astronomy.

\section{Application of analytical decomposition for describing trends in team evolution}

Analytical decomposition, introduced in the previous section, allows us to empirically derive the contribution of different modes of authorship over time and to explore the characteristics of teams as they evolve. We now fit Equation 1 to article teams in astronomy for all five-year time periods, from 1961 to 2010. Figure 6 (left panel) shows the change in the best-fit Poisson rates of both types of core teams as well as the evolution of the slope of the power-law component. As previously suggested, the Poisson rate of core teams has gradually increased from close to zero in the early 1960s to a little over three recently. On the other hand, the slope of the power-law component has gradually been flattening, from $\alpha = 6$ to $\alpha = 3$; i.e., the power-law component has been gaining in prominence. 

Figure 6 (middle panel) shows the relative contributions of the three modes of authorship in astronomy over the time period of 50 years, obtained by integrating the best-fit functional components. Remarkably, the contributions have remained relatively stable, with articles in the power-law component (i.e., articles produced by extended teams) making $\sim$30\%. This stability in the fraction of power-law articles is directly connected to the fixed propensity of authors to write articles with extended teams, as indicated in the model simulation. In all time periods most papers ($\sim$60\%) have been published by standard core teams (the Poisson component). Core teams with an extra member seem to appear in the early 1970s, but their contribution has remained at around 10\%.

\begin{figure*}
\centering
\includegraphics[width=1.0\textwidth]{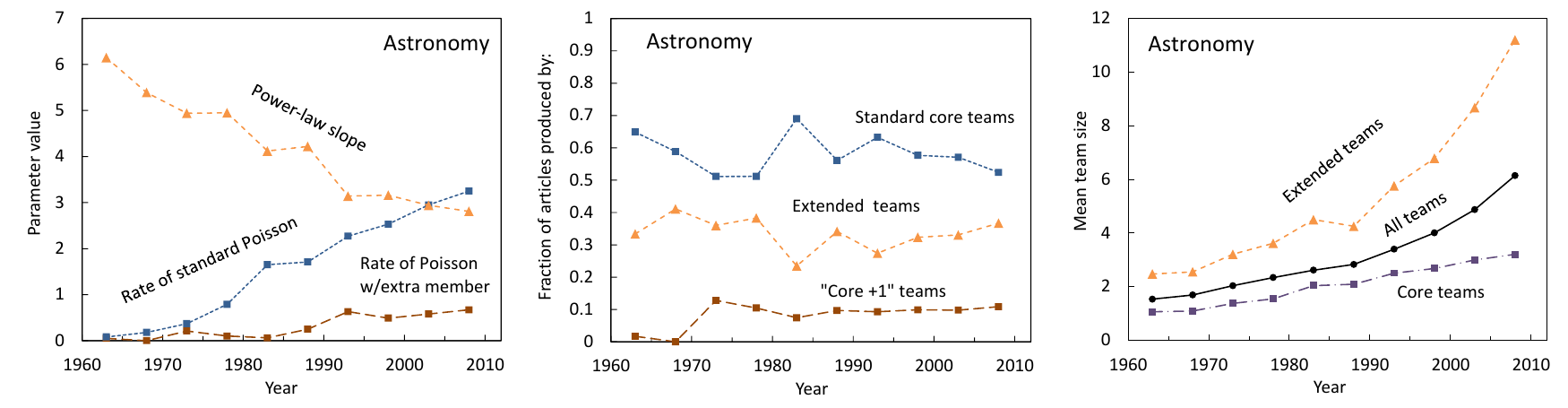}\caption{Trends in team evolution in astronomy from 1961-2010. Left: Fifty-year trend of parameters characterizing the three components of the distribution, derived from a functional fit (Equation 1). The characteristic size (i.e., Poisson rate) of standard core teams has ($\lambda_{P1}$) been rising throughout this period , while that of ``core +1" teams ($\lambda_{P2}$) has remained constant in the last two decades. The power-law slope ($\alpha_{PL}$) has been getting shallower, i.e. the significance of the power-law component has been increasing. Middle: Fraction of articles produced by different modes of authorship (team types): standard core ($f_{P1}$), ``core +1" ($f_{P2}$) and extended ($f_{PL}$). Right: Trends in the mean team size, overall ($\mu_{\rm all}$) and by team type (both types of core teams, $\mu  _{P}$, and extended teams, $\mu_{PL}$  . The increase in the overall mean team size in astronomy is primarily the result of the rapid growth of power-law (extended) teams.}
\label{fig:trends}
\end{figure*}

As pointed out earlier, many studies have emphasized the impressive growth of {\em mean} team sizes. We can now explore this trend in the light of the various authorship modes. In Figure 6 (right panel) we show the change in the mean size of all teams, and separately of core teams (standard and ``core +1" teams together) and of power-law (extended) teams. In the early 1960s both the core and the extended teams were relatively small (1.1 and 2.5 members, respectively). Subsequently, the mean size of core teams has increased linearly to 3.2 members. On the other hand, the mean size of extended teams has grown {\it exponentially}, and most recently averages 11.2 members. The exponential increase in the size of extended teams is affecting the overall mean, despite the fact that the extended teams represent the minority mode of authorship. While the growth of core teams is more modest, it nevertheless indicates that the level of collaboration, as measured by article team size, increases for this traditional mode of producing knowledge as well. Whether this increase is a reflection of a real change in the level of collaborative work or simply a change in the threshold for a contributor to be considered a coauthor is beyond the scope of this work.

In a similar fashion, we explored the evolution of fit parameters, mode contributions, and team sizes for mathematics and ecology (Figures S3 and S4). Mathematics features a small extended team component (10\%) that emerged in the mid-1980s. Extended teams in mathematics are still only slightly larger in size than the core teams. The share of ``core +1'' teams is increasing. The mean size of all core teams has increased, albeit moderately (from 1.2 to 1.8 members). In ecology the overall increase in mean team size mostly reflects the increase of the characteristic size of standard core teams in the 1980s. The observed increase of the share of extended teams appears to come at the expense of standard core teams. 

\section{Implications and conclusions}

The model proposed in this paper successfully explains the evolution of the sizes of scientific teams as manifested in author lists of research articles. It demonstrates that team formation is a multi-modal process. Primary mode leads to relatively small core teams the size of which may represent the typical number of researchers required to produce a research paper. The secondary mode results in teams that expand in size, and which are presumably employed to carry out research that requires expertise or resources from outside of the core team. These two modes are responsible for producing the hook and the power law-tail in team size distribution, respectively. 

This two-mode character may not be exclusive to team sizes. Interestingly, a similarly shaped distribution consisting of a hook and a power-law tail is characteristic of another bibliometric distribution, that of the number of citations that an article receives. Recently a model was proposed that successfully explained this distribution [33] by proposing the existence of two modes of citation, direct and indirect, where the latter is subject to cumulative advantage.

Understanding the distribution of the number of coauthors in a publication is of fundamental importance, as it is one of the most basic distributions that underpin our notions of scientific collaboration and the concept of ``team science''. The principles of team formation and evolution laid out in this work have the potential to illuminate many questions in the study of scientific collaboration and communication, and may have broader implications for research evaluation.

\begin{acknowledgments}
I thank the two anonymous reviewers for their constructive comments and Colleen Martin and John McCurley for copy editing.
\end{acknowledgments}

\end{article}

\end{document}